\documentclass[sigconf,nonacm]{acmart}
\pdfoutput=1

\usepackage{hyperref}
\usepackage{xspace}
\usepackage{listings}
\usepackage{subcaption}
\usepackage{tikz}
\usetikzlibrary{matrix}

\begin{document}

\title{No More Nulls!}
\author{Yisu Remy Wang}
\affiliation{%
  \institution{University of Washington}
  \city{}
  \country{}
}

\begin{abstract}
Since the inception of SQL, 
 nulls have frustrated database users and builders alike.
Those writing SQL must painstakingly
 guard their queries agaist surprising results caused by nulls,
 while those building database engines
 constantly struggle to implement the subtle semantics of 3-valued logic.
Given that the relational model already provides a way 
 to represent missing information,
 namely, with the absence of a tuple in a relation,
 one may step back and ask: ``Are nulls really necessary?''
We answer ``No!'' by proposing a new semantics for SQL
 that completely eliminates nulls.
Our semantics, called Columnar Semantics, 
 is as expressive as the standard 3-valued logic semantics,
 and behaves the same when the data and query are null-free.
Where the two semantics differ, Columnar Semantics results in simpler queries.
To evaluate Columnar Semantics and any other alternative semantics 
 or query languages, 
 we propose MIA (Missing Information Artifacts),
 a collection of queries and data sets for handling missing information,
 and invite contributions from the community.
\end{abstract}

\ccsdesc[500]{Information systems~Structured Query Language}

\keywords{SQL, nulls, missing information, 3 valued logic, normal forms} 

\maketitle

\section{Introduction}
Nulls in SQL are a pain.
Among many others,
 both the founder of the relational model, Codd~\cite{DBLP:books/aw/Codd90},
 and a co-inventor of SQL, Chamberlin~\cite{DBLP:conf/sigmod/Chamberlin23},
 have lamented the countless bugs caused by nulls,
 in both the database engine and the application code.
The current SQL standard supports nulls via 3-valued logic,
 which is a common source of confusion for developers~\cite{10.1145/3596673.3603142}.
Surprisingly, many attempts to address the problem of nulls
 have focused on increasingly complex many-valued logics~\cite{
  DBLP:conf/kr/ConsoleGL16,
  DBLP:journals/sigmod/Date08,
  DBLP:journals/sigmod/Gessert90,
  DBLP:conf/future/JiaFM92,
  10.1145/126482.126487} (all the way to 6-valued logic!).
These proposals have not seen wide adoption,
 because they are even harder to understand than 3-valued logic.
A recent work by Peterfreund and Libkin~\cite{DBLP:conf/pods/LibkinP23}
 goes the other direction, towards simplicity:
 they show that the textbook 2-valued logic suffices to capture 
 the semantics of SQL in the presence of nulls.
In this paper, we go a step further and argue that \textbf{nulls can be removed altogether}
 from the SQL language.
Our key insight is simple: nulls were invented to represent missing information,
 yet the relational model without nulls already provides a way to indicate information
 is missing, namely, with the absence of a tuple in a relation.
But what if only part of a tuple is missing?
Our solution is to decompose each relation into a collection of (correlated) columns,
 and an absent entry in a column thus represents missing information
 in the corresponding attribute of the tuple.
Specifically, \textbf{we propose Column Normal Form, 
 a new data normal form} inspired by Sixth Normal Form~\cite{DBLP:books/daglib/0014409} 
 and Graph Normal Form~\cite{RAIDocumentation}.
Based on Column Normal Form,
\textbf{we propose Columnar Semantics, a new semantics for SQL}
 where the query operates on a collection of columns instead of a collection of rows.

Column Normal Form improves upon the previous normal forms by allowing missing 
 information in any part of the tuple,
 even when the relation already satisfies no non-trivial dependencies.
Columnar Semantics satisfies the desiderata put forward by~\cite{DBLP:conf/pods/LibkinP23}:
 \begin{enumerate}
  \item It is as expressive as the standard 3-valued logic semantics.
  \item For null-free data and query, the behavior is identical to the standard semantics.
  \item When the two semantics differ, Columnar Semantics results in simpler queries.
 \end{enumerate}
While the first two criteria can be defined formally,
 the third one is rather subjective.
Peterfreund and Libkin~\cite{DBLP:conf/pods/LibkinP23} provides one interpretation 
 by measuring the size of the query.
We achieve simplicity by completely eliminating the complexity of nulls from all queries. 

The idea of handling nulls via normalization is not new.
LogicBlox and RelationalAI have built
 successful commercial databases based on Sixth Normal Form
 and Graph Normal Form~\cite{
  DBLP:books/daglib/0014409,
  RAIDocumentation,
  DBLP:conf/sigmod/ArefCGKOPVW15}.
In his keynote speech~\cite{DBLP:conf/sigmod/Chamberlin23},
 Chamberlin also pointed to normalization as 
 one of two candidate solutions 
 to the problem of missing information.
He also brought up the common criticism of normalization:
 decomposing the relations introduces additional joins,
 degrading query performance.
We follow a simple solution to this problem
 proposed by Peterfreund and Libkin~\cite{DBLP:conf/pods/LibkinP23}: 
 since our semantics is as expressive as the standard one,
 every query under our semantics can be {\em compiled} 
 into another query under the standard semantics,
 which is then executed by existing database engines.
In other words, we provide our null-free semantics as a
 ``front-end'', or ``user interface'', to the  programmer,
 while the ``back-end'' database engine remains unchanged.
On the other hand, the simpler semantics may also enable
 more sophisticated query optimization and execution techniques, 
 as evident in~\cite{DBLP:conf/pods/LibkinP23,DBLP:conf/sigmod/ArefCGKOPVW15,RAIDocumentation}.
In the future, more innovative systems can 
 directly implement Columnar Semantics 
 to take advantage of such opportunities.

The rest of this paper is organized as follows.
Section~\ref{sec:background} reviews background 
 on missing information in SQL and discusses related work.
Section~\ref{sec:cnf} and~\ref{sec:cs} introduces Column Normal Form 
 and Columnar Semantics, respectively.
Section~\ref{sec:cv3} compares Columnar Semantics with the standard 3-valued logic semantics, 
 and show them to be equally expressive.
Finally, Section~\ref{sec:conclusion}
 lays out future research directions and concludes.

\section{Background and Related Work}
\label{sec:background}

The history of nulls in SQL is almost as old as SQL itself,
 stretching back to the inception of the relational model some 50 years ago.
We therefore do not attempt to provide a comprehensive survey of the literature, 
 but rather focus on the fundamentals and the most relevant research.
In this section, we first review the standard semantics of SQL
 based on 3-valued logic.
Then, we discuss prior work on missing information that directly 
 inspired our approach,
 including data normalization and semantics based on 2-valued logic.

\subsection{SQL and 3-Valued Logic}
We present here a condensed overview of SQL and its semantics,
 aiming to provide intuition rather than completeness.
Our technique supports the full set of relational algebra 
 defined in~\cite{DBLP:conf/pods/LibkinP23}.

A {\em query} has the form:
\lstinline[mathescape=true]|SELECT $\vec{e},\vec{\gamma}$ FROM $\vec{R}$ WHERE $\phi$ GROUP BY $\vec{v}$|,
where $\vec{e}$ is a list of expressions,
 $\vec{\gamma}$ is a list of aggregate expressions,
 $\vec{R}$ is a list of relations,
 $\phi$ is a formula,
 and $\vec{v}$ is a list of attributes.
An {\em expression} is built up with functions 
 (like $+$ and $\times$),
 constants, and attributes;
 an {\em aggregate expression} is an
 aggregate funciton 
 (like \lstinline[mathescape=true]|COUNT| 
 and \lstinline[mathescape=true]|SUM|)
 applied to an expression;
 a {\em formula} consists of predicates over expressions
 (e.g., $e_1 = e_2$ and $e_1 < e_2$) 
 connected by the logical connectives 
 \lstinline|AND|,
 \lstinline|OR|, and
 \lstinline|NOT|.
We also support features like 
 \lstinline|IN|,
 \lstinline|ANY|, 
 \lstinline|ALL|,
 and subqueries, 
 by reducing them to the above constructs
 via standard techniques~\cite{DBLP:conf/pods/LibkinP23, DBLP:conf/btw/0001K15}.
 
Without aggregates, the meaning of a query can be understood
 as follows:
 we nondeterministically choose a tuple
 from each relation in $\vec{R}$ 
 and evaluate the formula $\phi$ over them;
 if $\phi$ evaluates to true,
 we evaluate each expression in $\vec{e}$ and output the result.
To evaluate aggregation,
 we first compute the tuples as above,
 then group them by the attributes in $\vec{v}$,
 and finally apply the aggregate function to each group.
When evaluating an expression or applying an aggregate function, 
 any null input results in a null output.
A predicate returns \lstinline|UNKNOWN| if 
 any of its arguments is null;
 one exception is the predicate ``\lstinline|IS NULL|''
 which returns \lstinline|TRUE| if its argument is null,
 and \lstinline|FALSE| otherwise.
In a formula, truth values are propagated
 following Kleene's 3-valued logic as follows, 
 where $t$ stands for \lstinline|TRUE|, 
 $f$ for \lstinline|FALSE|, and $u$ for \lstinline|UNKNOWN|:

\begin{tabular}{ c|ccc } 
  \lstinline|AND| & t & f & u \\ 
  \hline
  t & t & f & u \\
  f & f & f & f \\
  u & u & f & u
\end{tabular}\qquad
\begin{tabular}{ c|ccc } 
  \lstinline|OR| & t & f & u \\ 
  \hline
  t & t & t & t \\
  f & t & f & u \\
  u & t & u & u
\end{tabular}\qquad
\begin{tabular}{ c|c } 
  \lstinline|NOT| &  \\ 
  \hline
  t & f \\
  f & t \\
  u & u 
\end{tabular}\quad

\begin{example}
\label{ex:sql}
The 3-valued logic semantics can produce surprising results.
Consider the query \lstinline|SELECT * FROM R WHERE R.x = R.x|.
One would expect the result to be the same as the relation \lstinline|R|,
 due to the reflexivity of equality.
However, because \lstinline|=| returns \lstinline|UNKNOWN| 
 when either argument is null,
 the query drops all entries 
 where \lstinline|x| is null.
\end{example}

Surprises like Example~\ref{ex:sql} 
 can lead to subtle bugs in SQL queries. 
They also complicate the implementation of database engines, 
 as common algebraic properties like the reflexivity of equality
 can no longer be assumed when authoring optimization rules.

\subsection{Handling Nulls with 2-Valued Logic}

One source of confusion in 3-valued logic
 is the unfamiliar behavior of \lstinline|UNKNOWN|.
Most developers are accustomed to 2-valued logic,
 where the only truth values are \lstinline|TRUE| and \lstinline|FALSE|.
Fortunately, and perhaps surprisingly,
 2-valued logic can already capture the semantics of SQL 
 in the presence of nulls, 
 as shown by Peterfreund and Libkin~\cite{DBLP:conf/pods/LibkinP23}.
Their semantics requires only a simple modification:
 whenever an argument to a predicate is null,
 return \lstinline|FALSE| instead of \lstinline|UNKNOWN|.
This way, query evaluation may proceed purely in 2-valued logic
 following the standard behavior of the logical connectives.
Peterfreund and Libkin proved their new semantics 
 agrees with the standar 3-valued logic semantics
 when the data and query are null-free,
 and the two semantics have exactly the same expressive power.
Consequently, every query under the new semantics
 can be compiled into another query under the standard semantics,
 which is to be executed by an existing database engine.
In other words, adopting the new semantics 
 incurs no performance penalty,
 assuming a sufficiently sophisticated compiler.

Nevertheless, the new semantics is not without 
 controversy\footnote{The author of this paper was sandwiched between two sides of 
 a passionate debate during the Q\&A session of~\cite{DBLP:conf/pods/LibkinP23}.}.
For example, neither \lstinline|NULL = NULL| nor \lstinline|NULL $\neq$ NULL|
 is true, defying the intuition that \lstinline|NULL $\neq$ NULL|
 is the same as \lstinline|NOT (NULL = NULL)|.
Peterfreund and Libkin addresses this with
 a more complex interpretation of comparisons,
 but it's not clear if this alternative interpretation
 introduces further complications.
It appears our troubles will not end, 
 as long as nulls remain.

\subsection{Eliminating Nulls via Decomposition}
One alternative to nulls is to simply
 use the absence of a tuple to indicate missing information.
But when only part of a tuple is missing, 
 dropping the entire tuple loses information.
A solution is to decompose the relation into smaller ones
 such that each tuple is represented by the join 
 of smaller pieces, and each piece can be missing.
Usually, the relation is decomposed into a certain 
 {\em normal form}, 
 for example Sixth Normal Form~\cite{DBLP:books/daglib/0014409}
 and Graph Normal Form~\cite{RAIDocumentation}.
A relation is in normal form if it cannot be decomposed further,
This is the approach implemented by LogicBlox 
 and RelationalAI~\cite{RAIDocumentation,DBLP:conf/sigmod/ArefCGKOPVW15},
 and advocated by Date et.al.~\cite{DBLP:journals/sigmod/Date08,DBLP:books/daglib/0014409}.
We illustrate the idea with an example.



\begin{figure}
\begin{subfigure}{\linewidth}
\centering
\begin{tabular}{|c|c|c|}
\hline
\textsf{Author} & \textsf{Institute} & \textsf{Address} \\ 
\hline
Codd & IBM & San Jose \\
Chamberlin & IBM & \lstinline|NULL| \\
Boyce & \lstinline|NULL| & San Jose \\
\hline
\end{tabular}
\caption{Missing information is marked with \lstinline|NULL|.}
\label{fig:table-null}
\vspace{1em}
\end{subfigure}
\begin{subfigure}{\linewidth}
\centering
\begin{tabular}{|c|c|}
\hline
\textsf{Author} & \textsf{Institute} \\
\hline
Codd & IBM \\
Chamberlin & IBM \\
\hline
\end{tabular}
\hspace{1em}
\begin{tabular}{|c|c|}
\hline
\textsf{Author} & \textsf{Address} \\ 
\hline
Codd & San Jose \\
Boyce & San Jose \\
\hline
\end{tabular}
\caption{Decomposition into two null-free relations.}
\label{fig:table-normalized}
\end{subfigure}
\caption{A relation with nulls and its decomposition.}
\Description{A relation with nulls and its decomposition.}
\end{figure}

\begin{example}
\label{ex:normalization}
Consider the relation in Figure~\ref{fig:table-null}, 
 where two values are missing and marked with \lstinline|NULL|.
It can be decomposed into the two relations in Figure~\ref{fig:table-normalized}
 which hold the same information, but do not use nulls.
\end{example}

After decomposing every relation into ones without nulls,
 queries can be evaluated with the textbook 2-valued logic.
However, it is not always possible to fully eliminate nulls via decomposition.
In Example~\ref{ex:normalization}, 
 we have implicitely assumed the \textsf{Author} attribute 
 to be a key and may never be missing.
But if we add a tuple 
\begin{tabular}{|c|c|c|}
\hline
\lstinline|NULL| & UW & Seattle\\
\hline
\end{tabular}
 to the relation, 
 then the relation cannot be decomposed at all.
In other words, decomposition assumes at least one column must never contain nulls. 
Another common criticism of decomposition is that
 it introduces additional joins in each query, 
 which can degrade performance and readability.
For example, every join with the table in Figure~\ref{fig:table-null}
 becomes up to two joins with both tables in Figure~\ref{fig:table-normalized}.

We will see how the above shortcomings can be addressed 
 by Column Normal Form and Columnar Semantics.
Specifically, we allow missing information in any part of the relation;
 the query is as readable as the original one, and in many cases idential;
 and it is always possible to compile the query into the original one,
 losing no performance at all.

\section{Column Normal Form}
\label{sec:cnf}

We begin by addressing one limitation of the approach
 to missing information via decomposition,
 namely, the assumption that at least one column must never contain nulls.

Our solution is simple:
 for each input relation, 
 we introduce an additional {\em opaque} column 
 holding a primary key of the relation.
This column is opaque, meaning it may never be used 
 in the surface syntax of the query 
 (but we will use it in an intermediate syntax).
Intuitively, one may think of this column as holding 
 the ``row number'' of each tuple.
By definition, this new column guarantees
 that the relation always has a primary key that is never missing.
Furthermore, a relation with $k$ columns can be decomposed
 into $k+1$ relations:
 one relation has a single column holding the primary key,
 and the other $k$ relations 
 map each key to the corresponding column in the original relation.
We say these $k+1$ relations are in {\em Column Normal Form}.

\begin{figure}
\begin{subfigure}{0.09\linewidth}
\centering
\begin{tabular}{|c|}
\hline
\textsf{id} \\
\hline
1 \\
2 \\
3 \\
\hline
\end{tabular}
\end{subfigure}
\begin{subfigure}{0.31\linewidth}
\centering
\begin{tabular}{|c|c|}
\hline
\textsf{id} & \textsf{Author} \\
\hline
1 & Codd \\
2 & Chamberlin \\
3 & Boyce \\
\hline
\end{tabular}
\end{subfigure}
\centering
\begin{subfigure}{0.26\linewidth}
\centering
\begin{tabular}{|c|c|}
\hline
\textsf{id} & \textsf{Institute} \\
\hline
1 & IBM \\
2 & IBM \\
\hline
\end{tabular}
\end{subfigure}
\begin{subfigure}{0.26\linewidth}
\centering 
\begin{tabular}{|c|c|}
\hline
\textsf{id} & \textsf{Address} \\
\hline
1 & San Jose \\
3 & San Jose \\
\hline
\end{tabular}
\end{subfigure}
\caption{The relation in Figure~\ref{fig:table-null} in Column Normal Form.}
\Description{The relation in Figure~\ref{fig:table-null} in Column Normal Form.}
\label{fig:cnf}
\end{figure}

\begin{example}
\label{ex:cnf}
Figure~\ref{fig:cnf} shows the decomposition of the relation 
 in Figure~\ref{fig:table-null} into Column Normal Form.
Here we illustrate the opaque column with integer ids,
 but its data type does not matter as we never 
 materialize the relations during execution,
 as we will explain in Section~\ref{sec:cs}.
\end{example}

Decomposing into Column Normal Form allows us to
 represent missing information in any column,
 simply by dropping the entry in the corresponding relation.
We even allow an entire tuple to be missing --
 that is, a table row with empty cells (as opposed to the absence of a row).
In this case, we retain the entry in the relation holding only keys, 
 but drop the corresponding entries in all other relations.
Retainig empty rows is useful.
For example, five empty rows may indicate we lost the data
 of five correspondents in a survey, 
 whereas simply dropping the rows would lose this information.

We can already write queries directly over the relations in Column Normal Form. 
In principle, any query $Q$ against the original table 
 can be simulated by another query\footnote{Or another {\em set} of queries; more on this later.}
 $Q'$ against the normalized relations, 
 since the normalized relations contain the same information.
However, $Q'$ may be slower and more complex,
 because it must first join the normalized relations.
Explicitly joining together the normalized relations
 also requires writing down the opaque column
 in the query, further complicating the query.

\begin{figure}
\begin{subfigure}[t]{0.4\linewidth}
\begin{lstlisting}[showlines=true]
SELECT Address 
  FROM R 

 WHERE R.Author = "Codd"


\end{lstlisting}
\caption{Original SQL query.}
\label{fig:codd}
\end{subfigure}
\;
\begin{subfigure}[t]{0.5\linewidth}
\begin{lstlisting}
SELECT R_Adress.Address 
  FROM R_Author, R_Address,
       R_id
 WHERE R_id.id = R_Author.id 
   AND R_id.id = R_Address.id
   AND R1.Author = "Codd"
\end{lstlisting}
\caption{Simulating the query in~\ref{fig:codd}.}
\label{fig:codd-simulate}
\end{subfigure}
\caption{Simulating a SQL query over Column Normal Form.}
\Description{Simulating a SQL query over Column Normal Form.}
\label{fig:simulate}
\end{figure}

\begin{example}
\label{ex:cnf-query}  
The query in Figure~\ref{fig:codd},
 over relation \lstinline|R| in Figure~\ref{fig:table-null},
 can be simulated by the query in Figure~\ref{fig:codd-simulate},
 where \lstinline|R_id|, \lstinline|R_Author|, and \lstinline|R_Address|
 are the normalized relations for the key column, 
 the \textsf{Author} column, and the \textsf{Address} column, respectively.
In this example, the join with \lstinline|R_id| can be omitted, 
 but we keep it here to help explain the semantics in Section~\ref{sec:cs}.
\end{example}
To address the above problems, 
 we introduce a new semantic interpretation of the syntax of SQL,
 which will allow us to adapt any query to work over 
 the normalized relations with little or no change.
We call this new semantics {\em Columnar Semantics}.

\section{Columnar Semantics}
\label{sec:cs}

Intuitively, Columnar Semantics treats a standard SQL query 
 as a syntactice sugar for another set of {\em expanded queries}
 over the normalized relations.
The reason we need a {\em set} of queries is that
 the output of the original query under the 3-valued logic semantics may 
 also contain nulls, 
 therefore we must also represent the output relation in Column Normal Form. 
We stress that these decomposed relations,
 as well as the queries over them,
 are merely conceptual devices to help explain the semantics.
As every query under Columnar Semantics 
 is compiled to the standard semantics before it runs,
 the normalized relations are never materialized,
 and the queries over them are never executed.

\begin{figure}
\centering
\begin{tabular}{c}
\begin{lstlisting}
SELECT ids($\vec{R}$), rename$(e_1)$ FROM normalize$(\vec{R})$ 
 WHERE expand$(\phi)$ GROUP BY rename$(\vec{v})$
\end{lstlisting}\\
\begin{lstlisting}[basicstyle=\color{black!40}\ttfamily\small]
SELECT ids($\vec{R}$), rename$({\color{red} e_2})$ FROM normalize$(\vec{R})$ 
 WHERE expand$(\phi)$ GROUP BY rename$(\vec{v})$
\end{lstlisting}\\
\dots\\
\begin{lstlisting}[basicstyle=\color{black!40}\ttfamily\small]
SELECT ids($\vec{R}$), rename$({\color{red}\gamma_1})$ FROM normalize$(\vec{R})$ 
 WHERE expand$(\phi)$ GROUP BY rename$(\vec{v})$
\end{lstlisting}\\
\begin{lstlisting}[basicstyle=\color{black!40}\ttfamily\small]
SELECT ids($\vec{R}$), rename$({\color{red}\gamma_2})$ FROM normalize$(\vec{R})$ 
 WHERE expand$(\phi)$ GROUP BY rename$(\vec{v})$
\end{lstlisting}\\
\dots
\end{tabular}
\caption{Expanded queries of $Q$ in Definition~\ref{def:cs}.}
\Description{Expanded queries of $Q$ in Definition~\ref{def:cs}.}
\label{fig:expanded}
\end{figure}

\begin{figure}
\begin{align*}
\text{ids}(\vec{R}) &= \{ R_\text{id}.\text{id} \mid R \in \vec{R} \} \\
\text{rename}(e) &= e[R.a \mapsto R_a.a \text{ for all } R.a \in e] \\
\text{rename}(\gamma) &= \gamma[R.a \mapsto R_a.a \text{ for all } R.a \in \gamma] \\
\text{rename}(\vec{v}) &= \vec{v}[R.a \mapsto R_a.a \text{ for all } R.a \in \vec{v}] \\
\text{normalize}(R) &= \{ R_\text{id} \} \cup \{ R_a \mid a \in R \} \\
\text{normalize}(\vec{R}) &= \{ \text{normalize}(R) \mid R \in \vec{R} \} \\
\text{expand}(\phi) &= \phi[R.a \mapsto R_a.a \text{ for all } R.a \in \phi] \\
&\cup \{ R_a.\text{id} = R_\text{id}.\text{id} \mid R.a \in \phi \}
\end{align*}
\caption{Auxiliary functions used in Definition~\ref{def:cs}.}
\Description{Auxiliary functions used in Definition~\ref{def:cs}.}
\label{fig:aux}
\end{figure}

\begin{definition}[Expanded Queries]
\label{def:cs}  
Consider $Q$ of the form:
\begin{lstlisting}
SELECT $e_1, e_2, \ldots, \gamma_1, \gamma_2, \ldots$ FROM $\vec{R}$ WHERE $\phi$ GROUP BY $\vec{v}$
\end{lstlisting}
The set of {\em expanded queries} of $Q$ is shown in Figure~\ref{fig:expanded}.
The functions \lstinline|ids|, 
 \lstinline|rename|, 
 \lstinline|normalize|, and \lstinline|expand| 
 are defined in Figure~\ref{fig:aux}.
\lstinline|ids|($\vec{R}$) returns the set of opaque key columns
 of the normalized relations of $\vec{R}$.
Given an expression $e$, or an aggregate expression $\gamma$,
 or a list of variables $\vec{v}$, 
 \lstinline|rename| replaces each attribute $R.a$ 
 with the attribute $R_a.a$ 
 in the normalized relation $R_a$, 
 which corresponds to column $a$ in $R$.
\lstinline|normalize| replaces each relation $R$ 
 with the set of relations in $R$'s Column Normal Form.
Finally, \lstinline|expand| renames attributes as above, 
 and for each attribute $R.a$ in $\phi$, 
 introduces a new predicate $R_a.\text{id} = R_\text{id}.\text{id}$,
 where $R_\text{id}$ is the relation holding the opaque keys.
\end{definition}

\begin{example}
\label{ex:cs}
The expansion of the original query in Example~\ref{ex:cnf-query} is
 exactly the query that ``simulates'' it.
In other words, \textbf{we can keep the original query unchanged},
 and interpreting it under Columnar Semantics
 returns the same result!
Of course, the Columnar Semantics does not 
 always agree with the standard semantics
 in the presence of nulls.
\end{example}

One intuitive way to understand the Columnar Semantics 
 is to think of each relation as a collection of {\em correlated} columns.
Two entries in different columns are correlated, 
 when they belong to the same tuple in the original relation.
When we evaluate a query, the \lstinline|WHERE| clause 
 only evaluates to true when the values of all attributes
 in the same table are correlated.

There is a small problem with the current semantics.
Because we have eliminated all nulls from the data, 
 the predicate \lstinline|IS NULL| has lost its meaning.
Nevertheless, one may wish to find tuples with missing values,
 perhaps to fill in the missing data.
For example, the query 
\lstinline|SELECT Author FROM R WHERE Address IS NULL|
finds all authors with missing addresses.
To express the same query, 
 we add a new construct \lstinline|R MISSING a|
 to indicate a tuple in \lstinline|R|
 is missing the value of attribute \lstinline|a|.
During expansion, we desugar \lstinline|R MISSING a|
 to \lstinline|R$_\texttt{id}$.id NOT IN (SELECT id FROM R$_\texttt{a}$)|.

\section{Columnar Semantics Captures SQL}
\label{sec:cv3}

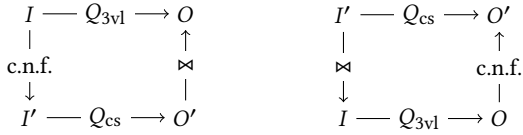
\begin{figure}
\begin{subfigure}{0.49\linewidth}
\centering
\begin{tikzpicture}
  \matrix (m) [matrix of math nodes, row sep=3em, column sep=5em]
  {
    I & O \\
    I' & O' \\
  };

  \draw[->] (m-1-1) -- (m-1-2) node[midway, align=center, fill=white] {$Q_\text{3vl}$};
  \draw[->] (m-1-1) -- (m-2-1) node[midway, align=center, fill=white] {c.n.f.};
  \draw[->] (m-2-1) -- (m-2-2) node[midway, align=center, fill=white] {$Q_\text{cs}$};
  \draw[->] (m-2-2) -- (m-1-2) node[midway, align=center, fill=white] {$\Join$};
\end{tikzpicture}
\end{subfigure}
\begin{subfigure}{0.49\linewidth}
\centering
\begin{tikzpicture}
  \matrix (m) [matrix of math nodes, row sep=3em, column sep=5em]
  {
    I' & O' \\
    I & O \\
  };

  \draw[->] (m-1-1) -- (m-1-2) node[midway, align=center, fill=white] {$Q_\text{cs}$};
  \draw[->] (m-1-1) -- (m-2-1) node[midway, align=center, fill=white] {$\Join$};
  \draw[->] (m-2-1) -- (m-2-2) node[midway, align=center, fill=white] {$Q_\text{3vl}$};
  \draw[->] (m-2-2) -- (m-1-2) node[midway, align=center, fill=white] {c.n.f.};
\end{tikzpicture}
\end{subfigure}
\caption{Relating 3-valued logic to Columnar Semantics.}
\Description{Relating 3-valued logic to Columnar Semantics.}
\label{fig:3vl-cs}
\end{figure}

Recall that the first two desiderata of any new semantics for SQL, 
 as proposed by Peterfreund and Libkin~\cite{DBLP:conf/pods/LibkinP23},
 require the new semantics to be as expressive as the standard one,
 and that they behave the same when the data and query are null-free.
These requirements ensure every SQL query can be expressed in the 
 new semantics, 
 and that we can always compile queries in the new semantics into the standard one,
 to be executed by existing database engines.
In this section, we show Columnar Semantics satisfies both requirements.
More precisely, we put forward the following proposition:
\begin{proposition}
\label{prop:cs-eq-3vl}
For every database instance $I$ and query $Q_\text{3vl}$,
 let $O$ be the result of running $Q_\text{3vl}$ on $I$, 
 under the standard 3-valued logic semantics.
Let $I'$ be the result of normalizing $I$ into Column Normal Form.
Then there is always a query $Q_\text{cs}$ such that
 running $Q_\text{cs}$ on $I'$ under Columnar Semantics returns $O'$,
 and the full outer join of $O'$ is the same as $O$.
And when $I$ and $Q_\text{3vl}$ are null-free, $Q_\text{cs}$ 
 is the same as $Q_\text{3vl}$.
\end{proposition}
\begin{proposition}
\label{prop:3vl-eq-cs}
Conversely, for every database instance $I'$ and query $Q_\text{cs}$,
 let $O'$ be the result of running $Q_\text{cs}$ on $I'$, 
 under Columnar Semantics.
Let $I$ be the full outer join of $I'$.
Then there is always a query $Q_\text{3vl}$ such that
 running $Q_\text{3vl}$ on $I$ under the standard 3-valued logic semantics returns $O$,
 and normalizing $O$ into Column Normal Form returns $O'$.
When $I'$ does not contain missing entries and $Q'$ does not use {\em \lstinline|MISSING|},
 $Q_\text{3vl}$ is the same as $Q_\text{cs}$.
\end{proposition}
Figure~\ref{fig:3vl-cs} illustrates the propositions, 
 where ``c.n.f.'' means normalizing each table in $I$ into Column Normal Form,
 and $\Join$ takes the full outer join of every group of normalized tables.

Instead of a formal proof, 
 we provide an intuitive argument based on the following observation:
 if we equate a missing entry with a null,
 then \textbf{Columnar Semantics can simulate the 2-valued logic semantics} 
 by Peterfreund and Libkin~\cite{DBLP:conf/pods/LibkinP23}.
Reusing their result that the 2-valued logic semantics
 captures the standard 3-valued logic semantics,
 we may conclude the Columnar Semantics also captures the standard semantics.

First, when the data and query are null-free,
 Columnar Semantics coincides with the standard semantics.
This is because decomposing to Column Normal Form 
 is lossless, and the expanded queries 
 join the normalized relations back together 
 into the original relation.

Let us now consider the evaluation of expressions in the presence of nulls.
In both the standard semantics and the 2-valued logic semantics,
 an expression evaluates to null if any of its arguments is null.
In Columnar Semantics, 
 a null corresponds to a missing value, 
 and an expression over missing values 
 will not be evaluated in the first place,
 therefore it also ``produces a missing value''.
This is equivalent to producing a null in the standard semantics.
The same reasoning also applies to the evaluation of aggregates.

Next we focus on the formula $\phi$ in the \lstinline|WHERE| clause.
Both the 2-valued logic semantics and Columnar Semantics
 propagate truth values according to the standard 
 truth tables of the logical connectives.
The only difference is the behavior of predicates over missing values.
In the 2-valued logic semantics, 
 a predicate always returns \lstinline|FALSE| when any of its arguments is null.
In Columnar Semantics, once again, 
 the predicate would not be evaluted over missing values in the first place.
In the absence of negations,
 both behaviors result in no output and are equivalent.
When $\phi$ contains negations, however, 
 the two semantics may differ.
\begin{example}
Consider the following query:
\begin{lstlisting}
 SELECT * FROM R WHERE NOT (R.x = R.x)
\end{lstlisting}
Under the 2-valued logic semantics, 
 the query returns all tuples where \lstinline|x| is null, 
 because \lstinline|NOT (NULL = NULL)| is true.
Under Columnar Semantics, 
 the query returns nothing, 
 because the predicate \lstinline|R.x = R.x| is not evaluated 
 over missing values.
\end{example}
We would argue that the result produced 
 by Columnar Semantics is more intuitive\footnote{On this example, 
 Columnar Semantics agrees with 3-valued logic semantics.};
 but since our goal is to capture the 2-valued logic semantics,
 we must find a way to simulate its behavior.
Our solution is as follows.
We first push down all negations to the leaves of the formula, 
 by applying De Morgan's laws.
Then, we replace negated predicate \lstinline|NOT P(R.x, S.y)|
 with the following:
\begin{lstlisting}
(R MISSING x) OR (S MISSING y) OR NOT P(R.x, S.y)
\end{lstlisting}
This simulates the behavior of \lstinline|NOT P(R.x, S.y)|
 under the 2-valued logic semantics, 
 where it returns true when either argument is null.

Our transformation above ensures the resulting query 
 has size linear in the size of the query in 2-valued logic.
Together with the result by Peterfreund and Libkin~\cite{DBLP:conf/pods/LibkinP23}
 that the standard SQL semantics can be simulated by 
 a linear-size query in 2-valued logic semantics,
 we conclude that Columnar Semantics 
 can capture the standard semantics in linear size as well.


\section{Discussion}
\label{sec:conclusion}

With Column Normal Form and Columnar Semantics, 
 we take the first step towards a semantics for SQL
 completely without nulls.
Yet much remains to be done.
One immediate task is to formally prove our claims
 in Propositions~\ref{prop:3vl-eq-cs} and~\ref{prop:cs-eq-3vl}.
But even if we are certain that Columnar Semantics 
 captures the standard semantics,
 it is far from obvious that 
 we can always compile every Columnar Semantics query 
 into an {\em optimal} query in the standard semantics.
An effective optimizer is necessary to make 
 Columnar Semantics practical.
Another real-world concern is that,
 adopting Columnar Semantics 
 as a front-end for SQL introduces an additional abstraction layer,
 which can complicate debugging and profiling.

A potential solution to the challenges in 
 both performance and the user experience
 is to directly execute Columnar Semantics
 without the detour to 3-valued logic.
As a side effect of its design, the Columnar Semantics
 may in fact be more amenable to modern query processing techniques.
For example, algorithms like Worst-Case Optimal Joins~\cite{DBLP:journals/jacm/NgoPRR18}
 can run {\em faster} when there are more, skinnier tables. 
Relations in Column Normal Form can also be considered 
 as a special case of factorized databases~\cite{DBLP:journals/sigmod/OlteanuS16}, 
 so factorized query processing techniques may apply.
Last but not least, 
 it is no coincidence that the names ``Column Normal Form'' and ``Columnar Semantics''
 are reminiscent of the columnar architecture for databases.
As such architecture already stores tables by columns, 
 Columnar Semantics may be even ``closer to the metal'' 
 than the standard semantics.

We have focused on the comparison of Columnar Semantics 
 with the standard 3-valued logic semantics using nulls.
But there are already successful databases
 that do not use nulls at all,
 as implemented by LogicBlox and RelationalAI~\cite{RAIDocumentation,DBLP:conf/sigmod/ArefCGKOPVW15}.
What is our contribution, then?
We think of the approach of these existing systems
 as ``manual null management'', 
 where the developer explicitly specifies what to do with missing information.
This is beneficial in some cases where one requires
 fine-grained control over the behavior of missing data.
Our approach can be thought of as ``automatic null management'',
 where we automatically normalize the relations 
 and represent each missing value with an absent entry.
Indeed, loading a \textsf{CSV} file to RelationalAI 
 will create a relation where each row is assigned a row number, 
 and the first action performed by the user is almost always 
 to manurally decompose this relation into Graph Normal Form~\cite{RAIDocumentation}.

By the end of the day, \textbf{the only salient way to test a semantics for SQL
 is to write queries with it}.
Can the semantics help analysts ask real questions 
 over real data?
Is it more clear than SQL, 
 and can it be efficiently implemented by a database engine?
To answer these questions, 
 we propose MIA\footnote{MIA is available at \url{https://github.com/remysucre/mia/}} (Missing Information Artifacts),
 a collection of queries and datasets 
 for handling missing information.
We initialized the collection with the data release 
 from the SQLShare project~\cite{DBLP:conf/sigmod/JainMHHL16},
 containing thousands of queries written to analyze scientific data.
MIA serves as a test bed for new semantics for SQL, 
 or even entirely new query languages:
 if you believe your semantics/language is better than SQL,
 try rewriting the queries in MIA with it!

We conclude by noting that
 the Column Normal Form has the potential of 
 resolving the ``second greatest sin'' of SQL, 
 namely, allowing duplicate rows in a relation.
Again, both Codd and Chamberlin have argued 
 vigorously against this design~\cite{DBLP:books/aw/Codd90,DBLP:conf/sigmod/Chamberlin23}.
In Column Normal Form, 
 duplicate rows are impossible by definition,
 because every row has a unique opaque key.
However, we shall defer further discussions 
 to the future -- starting one fire is enough.

\begin{acks}
We thank Liat Peterfreund, Leonid Libkin, 
 Stanley Yang, and Dan Grossman for stimulating 
 discussions on the topic of this paper.
\end{acks}

\bibliographystyle{ACM-Reference-Format}
\bibliography{nulls}

\end{document}